\documentclass[conference]{IEEEtran}
\IEEEoverridecommandlockouts
\usepackage{balance}
\usepackage{cite}
\usepackage{amsmath,amssymb,amsfonts}
\usepackage{algorithmic}
\usepackage{graphicx}
\usepackage{textcomp}
\usepackage{xcolor}
\usepackage{url}
\usepackage[hidelinks]{hyperref}
\usepackage[T1]{fontenc}

\newcommand{\citeneeded}[1]{\textcolor{red}{[cite needed]}}
\usepackage{soul}

\definecolor{greenBello}{HTML}{D9EAD3}
\definecolor{redBello}{HTML}{F4CCCC}

\usepackage{cellspace, tabularx}
\addparagraphcolumntypes{X}
\renewcommand\tabularxcolumn[1]{m{#1}}

\usepackage{colortbl}
\usepackage{framed}
\usepackage[moderate, tracking=normal, leading=normal]{savetrees}
\usepackage{enumitem}
\usepackage{scalerel}
\usepackage{changepage}

\definecolor{formalshade}{rgb}{0.9,0.95,1}

\definecolor{darkblue}{rgb}{0.6,0.75,0.95}

\newenvironment{formal}{%
  \MakeFramed{\advance\hsize-\width\FrameRestore}%
  \noindent\hspace{-4.55pt}%
  \begin{adjustwidth}{}{7pt}%
}
{%
  \end{adjustwidth}\endMakeFramed%
}

\definecolor{formalshade2}{rgb}{1,0.85,0.85}
\definecolor{darkred}{rgb}{0.6,0.0,0.0} %

\usepackage{listings}
\usepackage{csquotes}
\usepackage{multirow}
\usepackage{comment}
\usepackage{tabularx}
\usepackage[most]{tcolorbox}
\newtcolorbox{LLMsnippet}{enhanced jigsaw,breakable, sharp corners,  colback=white,colframe=black, size=title, boxrule=0.5pt}
\newcommand{\addr}[1]{{\fontfamily{cmtt}\selectfont{#1}}}

\usepackage[acronym, shortcuts]{glossaries-extra}

\glssetcategoryattribute{acronym}{nohyper}{true}
\setabbreviationstyle[acronym]{long-short}

\newacronym{ai}{AI}{Artificial Intelligence}
\newacronym{dl}{DL}{Deep Learning}
\newacronym{llm}{LLM}{Large Language Model}
\newacronym{ml}{ML}{Machine Learning}
\newacronym{rag}{RAG}{Retrieval Augmented Generation}
\newacronym{sysadmin}{SysAdmin}{System Administrator}
\newacronym{cot}{CoT}{Chain-of-Thought}

\newcommand{\netOne}{\emph{WebServer}}
\newcommand{\netTwo}{\emph{Routers}}
\newcommand{\netThree}{\emph{Intradomain}}

\def\BibTeX{{\rm B\kern-.05em{\sc i\kern-.025em b}\kern-.08em
    T\kern-.1667em\lower.7ex\hbox{E}\kern-.125emX}}
\begin{document}

\title{Can LLMs Understand Computer Networks? Towards a Virtual System Administrator}
\IEEEpeerreviewmaketitle

\author{\IEEEauthorblockN{Denis Donadel\IEEEauthorrefmark{1},
Francesco Marchiori\IEEEauthorrefmark{1},
Luca Pajola\IEEEauthorrefmark{1}\IEEEauthorrefmark{2},
Mauro Conti\IEEEauthorrefmark{1}\IEEEauthorrefmark{3}}
\IEEEauthorblockA{\IEEEauthorrefmark{1}Department of Mathematics\\
University of Padova,
Padua, Italy}
\IEEEauthorblockA{\IEEEauthorrefmark{2}Spritz Matter Srl, \\Padua, Italy}
\IEEEauthorblockA{\IEEEauthorrefmark{3}Faculty of Electrical Engineering, Mathematics and Computer Science\\
Delft University of Technology, Delft, The Netherlands\\\textit{denis.donadel@phd.unipd.it, francesco.marchiori.4@phd.unipd.it, luca.pajola@spritzmatter.com, mauro.conti@unipd.it}}}

\maketitle

\begin{abstract}
Recent advancements in Artificial Intelligence, and particularly Large Language Models (LLMs), offer promising prospects for aiding system administrators in managing the complexity of modern networks.
However, despite this potential, a significant gap exists in the literature regarding the extent to which LLMs can understand computer networks.
Without empirical evidence, system administrators might rely on these models without assurance of their efficacy in performing network-related tasks accurately.

In this paper, we are the first to conduct an exhaustive study on LLMs' comprehension of computer networks.
We formulate several research questions to determine whether LLMs can provide correct answers when supplied with a network topology and questions on it.
To assess them, we developed a thorough framework for evaluating LLMs' capabilities in various network-related tasks.
We evaluate our framework on multiple computer networks employing proprietary (e.g., GPT4) and open-source (e.g., Llama2) models.
Our findings in general purpose LLMs using a zero-shot scenario demonstrate promising results, with the best model achieving an average accuracy of 79.3\%.
Proprietary LLMs achieve noteworthy results in small and medium networks, while challenges persist in comprehending complex network topologies, particularly for open-source models.
Moreover, we provide insight into how prompt engineering can enhance the accuracy of some tasks.
\end{abstract}


\section{Introduction}
\label{sec:introduction}

The increasing complexity of computer networks raises the need for technical experts to maintain the operation, configuration, and security of computer systems and networks within an organization's perimeter. \acp{sysadmin} are generally the professionals designed to take care of these tasks, performing various tasks such as router and switch configuration, resolving connectivity issues, and even analyzing intrusion detection attempts~\cite{sommestad2013intrusion}.
They are responsible for network design, implementation, configuration, and maintenance, and they need to provide security by constant monitoring and troubleshooting.
However, expanding network complexity and the proliferation of cyber threats make their job increasingly convoluted.
The integration of \ac{ai} powered assistants has emerged as a potential solution.
In particular, the recent advancements in the field of \acp{llm} have made their usage as an assistant to human operation increasingly popular~\cite{ross2023programmer}.
Furthermore, companies and organizations may not have the resources to maintain their IT infrastructure, and, as such, virtual \acp{sysadmin} can significantly help automatically identify problems and vulnerabilities.  %

A \ac{llm} is a type of \ac{dl} model designed to generate human-like text based on the input it receives.
These models are trained on vast amounts of text data to learn patterns, structures, and relationships in language.
This knowledge allows them to solve diverse tasks (e.g., translation, summarization, question answering, text and image generation), contrasting with prior models confined to solving specific tasks~\cite{chang2023survey}.
This has been possible because of the improved reasoning and generalization capabilities with respect to \ac{ml} models~\cite{wei2022chain, huang2022towards}. 
Indeed, \ac{ml} and \ac{dl} have found applicability in networking, helping on various tasks~\cite{wang2017machine, zhang2019deep}.
As such, using \ac{ai} in computer networks is not new.
However, up to date, while some works have investigated the usage of \acp{llm} in the context of computer networks, none of them have studied their understanding of the topology and their possible usage as assistants for \acp{sysadmin}~\cite{wang2023network, jiang2023large}.

This research investigates how and how much \acp{llm} can integrate the two aspects and provide correct answers on computer networks presented as graphs.
Such capabilities can be fundamental in different sections.
Network engineering can benefit from \acp{llm} as experienced peers when designing or updating network topologies.
Moreover, they can help in spotting erorrs inside networks during troubleshooting.
Security engineers can also benefit from \acp{llm} capable of understanding network topologies.
They can query \acp{llm} to identify security issues and misconfigurations faster.

With respect to pure graph networks, computer networks implicitly include certain information masqueraded into other details.
Links between nodes provide a clear example.
Two nodes exposing IP addresses belonging to the same subnetwork can be represented with a direct link, even if it is not explicitly described in the graph formalization.
On the other hand, computer networks generally do not include self-loops or direct edges, which can simplify the analysis by reducing the search space. 

\textit{Contributions.}
This paper investigates the issue of \ac{llm} understanding of computer networks by proposing a framework for their evaluation.
With the selection of six state-of-the-art \acp{llm} and the representation of different networks, we define four research questions to assess the capabilities of these models.
Our results indicate that \acp{llm} can comprehend basic network structures and help administrators in several tasks.
However, they still struggle in specific scenarios where the network topology complexity increases.
Furthermore, we conduct a study on prompt engineering and provide insightful takeaways for getting the best performance from \acp{llm}.
Our contributions can be summarized as follows.
\begin{itemize}
    \item We define a framework to assess the capabilities of \acp{llm} in understanding computer network topologies.
    \item We conduct experiments on state-of-the-art models (3 proprietary and 3 open-source) over 3 different network topologies of increasing complexity.
    \item We discuss the results, providing insight into possible prompt engineering techniques to improve results. 
    \item We open-source our networks and prompts at: \url{https://github.com/spritz-group/LLM-SysAdmin}.
\end{itemize}

\textit{Organization.}
The rest of the paper is organized as follows.
Section~\ref{sec:methodology} explains research questions and our methodology.
In Section~\ref{sec:experiments}, we provide details on our experiments, the results of which are shown in Section~\ref{sec:results}.
We discuss our results in Section~\ref{sec:discussion}, and provide related works in the literature in Section~\ref{sec:related}.
Finally, Section~\ref{sec:conclusions} concludes our work.

\section{Methodology}\label{sec:methodology}

\textit{Problem Statement.} A preliminary question needs to be addressed to provide the basis for addressing the discussed problems: \emph{are \acp{llm} able to comprehend the structures of computer networks?}
Specifically, this paper addresses the problem by investigating the following research questions.

\begin{itemize}
    \item \textbf{RQ1}: Are \acp{llm} able to answer basic questions on network topologies correctly?
    \item \textbf{RQ2}: Can \acp{llm} provide graphical representations of network topologies? 
    \item \textbf{RQ3}: Can \acp{llm} recognize subnetworks and IP addresses inside computer networks?
    \item \textbf{RQ4}: Can \acp{llm} comprehend computer network connections?
\end{itemize}

We selected this set of research questions as they most closely reflect the capabilities a \ac{sysadmin} assistant must have.
As such, through answering our research questions, we investigate the concept of \textit{\acp{llm} comprehension of computer networks} and allow us to objectively evaluate the utility of \acp{llm} in this context.

\begin{figure}[htb]
    \centering
    \includegraphics[width=\columnwidth]{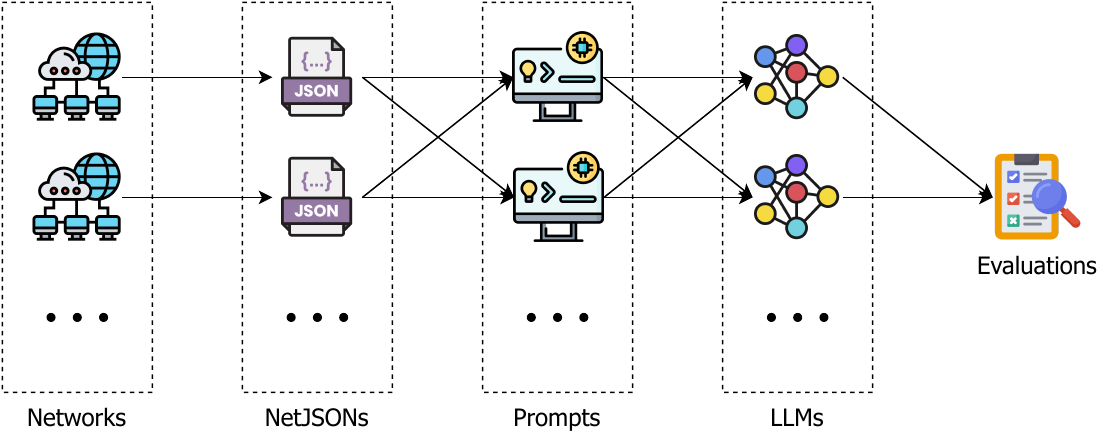}
    \caption{Our evaluation framework. We test several \acp{llm} with a combination of tasks specific to each network which is represented in a NetJSON format.}
    \label{fig:framework}
\end{figure}

\textit{Framework.}
To the best of our knowledge, no frameworks or previous work have analyzed the problem of computer network understanding on \acp{llm}.
To test their capabilities, we define a list of questions whose response is closely related to a specific network architecture.
As such, we devise queries that cannot be answered without a certain degree of knowledge of the underlying network structure.
We represent our general framework in Figure~\ref{fig:framework}.

Using in-context learning, we include the network graph directly in the prompt. Other solutions, such as \ac{rag}~\cite{lewis2020retrieval}, may be used for more extensive networks that will not fit the prompt size.
However, since modern \acp{llm} have quite big context sizes, it is usually enough for the scope of this work.
One example is ClaudeAI, which supports a 100k tokens context-window~\cite{claudeai}.
Moreover, we only test a zero-shot prompt style for several reasons.
First, being this research the first on the topic, we decide to start with the most accessible and most straightforward approach.
Moreover, a few-shot prompt may be infeasible to be developed in a real-world application because of the variety of possible questions and tasks~\cite{wang2020generalizing}.  

\textit{Prompting.}
The straightforward primary prompt we use as a template is shown in Figure~\ref{fig:prompt}.
Based on the model used, we adapted the template to suit the model's need, e.g., adding keywords to specify where the instruction starts and ends.
The prompt begins with a sentence introducing the context of the questions.
Then, a fundamental issue regards the representation of the network graph.

\begin{figure}[hb]
\begin{LLMsnippet}
\footnotesize
{
\fontfamily{qcr}\selectfont{
To answer the following questions, consider the following network:\\
\`{}\`{}\`{}\\
\{network\}\\
\`{}\`{}\`{}\\
\\
\{task\}
}
} 
\end{LLMsnippet}
\caption{Basic prompt for network comprehension task.}
\label{fig:prompt}
\end{figure}

\textit{Network Representation.}
To date, there is no widely accepted standardized way to encode graphs and basic information of a computer network in a textual-based format.
Some techniques have been developed regarding general graphs, from essential solutions like adjacency lists or edge lists to more feature-rich Graph Modelling Language~\cite{himsolt1997gml} and Graph Markup Language~\cite{brandes2013graph}.
However, since our network has a precise meaning and these languages are general-purpose, we decided to employ a standard designed ad-hoc for computer networks.
The most widely adopted solution is NetJSON~\cite{netjson}, which we employ in our experiments.
Following this network structure, we insert each instruction we want to evaluate our model on one by one.
The employed networks are available on our Github repository\footnote{\label{fn:git}\url{https://github.com/spritz-group/LLM-SysAdmin}}.

\section{Experiments}
\label{sec:experiments}

\textit{Models.}
As the first research study, we investigate freely accessible models (or subscription-based, avoid per-token payments), which can represent a baseline for further studies.
Therefore, we choose three different off-the-shelf online \acp{llm}, namely:
\begin{itemize}
    \item Bing Copilot (GPT-4 based, 1756B parameters)~\cite{bing},
    \item Github Copilot (GPT-4 based, 1756B parameters)~\cite{github},
    \item Chat-GPT (GPT-3.5 based, 175B parameters)~\cite{chatgpt}.
\end{itemize}
Moreover, we were interested in smaller models that could also run locally.
Therefore, we locally run some open-source models and test the same queries on them: 
\begin{itemize}
    \item Llama 2 (13B parameters)~\cite{touvron2023llama},
    \item Mistral Instruct v0.2 (7B parameters)~\cite{jiang2023mistral},
    \item BASH Coder Mistral (7B parameters) merged with Mistral Instruct v0.2 (7B parameters) using slerp~\cite{bash,slerp}.
\end{itemize}
While online and proprietary models are generally more powerful~\cite{gpt4research, sakaguchi2019winogrande}, local models provide enhanced flexibility, are not subject to API rate limits, and are not required to pay a fee for each token as pay-per-use models.
Moreover, computer network structures may include sensible information or can be used to infer security defences~\cite{la2014role}.
This may discourage safety-critical actors from sharing such information with services owned by other companies so as not to risk compromising their privacy. This topic is discussed more into detail in Section~\ref{subsec:privacy}.

\textit{Tasks.}
To understand the level of knowledge a model can get regarding network structures, we design a list of tasks to grasp the \acp{llm} comprehension of various aspects.
We show each task in Table~\ref{tab:quest_gen}.
In particular, we took inspiration from computer network books~\cite{tanenbaum2003computer} and exercise lists~\cite{ex-usi}.
We engineer the queries through several tests to ensure that the understanding of each question is clear enough not to be misinterpreted by the majority of the models and by humans.
However, we avoid excessive tuning on the question format at this stage, trying to present questions in a direct form that could be employed by a system administrator seeking help in a \ac{llm} to solve a network issue.
We discuss possible prompt engineering improvements in Section~\ref{subsec:prompt_eng}.
Up to \hyperref[tab:quest_gen]{\texttt{T9}}, tasks are the same for each scenario.
From \hyperref[tab:quest_gen]{\texttt{T10}} on, questions require two node names to compute subnetworks and connections between them.
In particular, \hyperref[tab:quest_gen]{\texttt{T11}} looks for direct connections between nodes that are \emph{not} directly connected, while \hyperref[tab:quest_gen]{\texttt{T12}} expects a positive answer.
The specific nodes we employed are available in our Github repository\textsuperscript{\ref{fn:git}}.
Finally, the \acp{llm}' replies to each task are evaluated by computer science PhD students as either correct or wrong.
Since the networks' complexity level is quite low, only basic topology knowledge is required to evaluate these questions.

\renewcommand\tabularxcolumn[1]{m{#1}}
\newcolumntype{C}{>{\centering\arraybackslash}X}

\begin{table}[b]
\caption{Tasks used for each RQ. From task \texttt{T10} on, each question has different \texttt{x1} and \texttt{x2} based on the network.}
\label{tab:quest_gen}
\centering
\begin{tabularx}{\columnwidth}{>{\hsize=0.075\hsize}C|>{\hsize=0.075\hsize}C|X} \hline 
\textbf{RQ} & \textbf{ID}  & \textbf{Task}                                                                                             \\ \hline
\multirow{3}{*}{\rotatebox{90}{\begin{tabular}[c]{@{}c@{}}\textbf{RQ1}\\(Topol.)\end{tabular}}}    & \cellcolor{gray!15} \texttt{T1}  & \cellcolor{gray!15} How many nodes are there in the network? Answer with a number.                                   \\
                        & \texttt{T2} & How many IP addresses are assigned to devices?                              \\
                        & \cellcolor{gray!15} \texttt{T3}  & \cellcolor{gray!15} Which devices have the most IP addresses assigned?             \\ \hline
\rotatebox{90}{\begin{tabular}[c]{@{}c@{}}\textbf{RQ2}\\(Draw)\:\end{tabular}}
& \texttt{T4}  &  Draw me the graph of my network. If you can't draw it, use ascii art.                                      \\ \hline
\multirow{9}{*}{\rotatebox{90}{\begin{tabular}[c]{@{}c@{}}\textbf{RQ3}\\(Addressing)\end{tabular}}}
                        & \cellcolor{gray!15} \texttt{T5} & \cellcolor{gray!15} Are there any devices with special-purpose IP addresses (e.g., multicast addresses)? \\
                        & \texttt{T6}  & Do any devices have multiple IP addresses assigned to them?     \\
                        & \cellcolor{gray!15} \texttt{T7}  & \cellcolor{gray!15} Are there any IPv6 addresses assigned?   \\

                        & \texttt{T8}  & How many subnetworks are there in my network? Answer with a number.                             \\
                        & \cellcolor{gray!15} \texttt{T9}  & \cellcolor{gray!15} Is it possible to remove one subnetwork but keeping all the devices able to ping each other?   \\
                        & \texttt{T10} & Which is the subnetwork that connects x1 to x2? \\ 
                        \hline
\multirow{4}{*}{\rotatebox{90}{\begin{tabular}[c]{@{}c@{}}\textbf{RQ4}\\(Paths)\end{tabular}}}    & \cellcolor{gray!15} \texttt{T11} & \cellcolor{gray!15} Is it possible for x1 to ping x2 without any hop? Answer directly with "yes" or "no". (negative) \\
                        & \texttt{T12} & Is it possible for x1 to ping x2 without any hop? Answer directly with "yes" or "no". (positive) \\
                        & \cellcolor{gray!15} \texttt{T13} & \cellcolor{gray!15} Traceroute from x1 to x2.                                                                        \\ \hline
\multicolumn{3}{l}{\footnotesize{\texttt{T11} includes only cases in which x1 cannot ping x2 without any hops.}} %
\end{tabularx}
\end{table}

\textit{Networks.} We extract three increasingly complex networks from Kathara~\cite{scazzariello2020kathara} main labs, and we convert each in the NetJSON format~\cite{netjson}.
We edited addresses and links in some labs to create a more challenging environment.
Furthermore, this procedure ensures that \acp{llm} will deal with unforeseen data, as their training datasets are not accessible but might include the original Kathara labs~\cite{scazzariello2020kathara}.
The most straightforward setup comprises two devices only: a client and a \netOne{} (denoted \textbf{W} in the results) exposing an example webpage.
Both devices expose a link-local IPv6 address and are connected to the same IPv4 subnetwork.
Then, we consider a slightly more complicated network including three \netTwo{} and five subnetworks connecting them. This network is denoted as \textbf{R} in the results.
Finally, the most complex architecture includes 12 nodes interconnected through 15 different subnetworks.
This last scenario represents an \netThree{} routing situation, denoted \textbf{I} in the results.

\textit{Evaluation.}
Most questions accept a single precise answer that cannot leave space for interpretation.
We assign 1 point to the \ac{llm} for correct and complete answers and 0 for incorrect questions.
Moreover, correct but incomplete answers are awarded with 0.5 points.
An example is question  \hyperref[tab:quest_gen]{\texttt{T9}}  \emph{``Which devices have the most IP addresses assigned?''}: if more than one device has the same number of addresses, which is also the maximum, returning only one of their IDs is considered partially correct but incomplete.
Another particular example is the drawing of the network graph, where we graded incomplete but still meaningful answers with 0.5 points.  
Since LLM answers are not deterministic, each networking question is repeated and evaluated 10 times.

\section{Results}\label{sec:results}

We applied our newly developed framework on the 6 \acp{llm} described, conducting ten tests for each task and each network. Therefore, a score of 10 means that all the tasks obtained a correct answer. Conversely, 0 indicates that wrong answers only. In this section, we present the overall results and address the research questions we formulated initially.

\subsection{Overall results}\label{subsec:overall_res}
\begin{figure}[htb]
    \centering
    \includegraphics[width=.9\columnwidth]{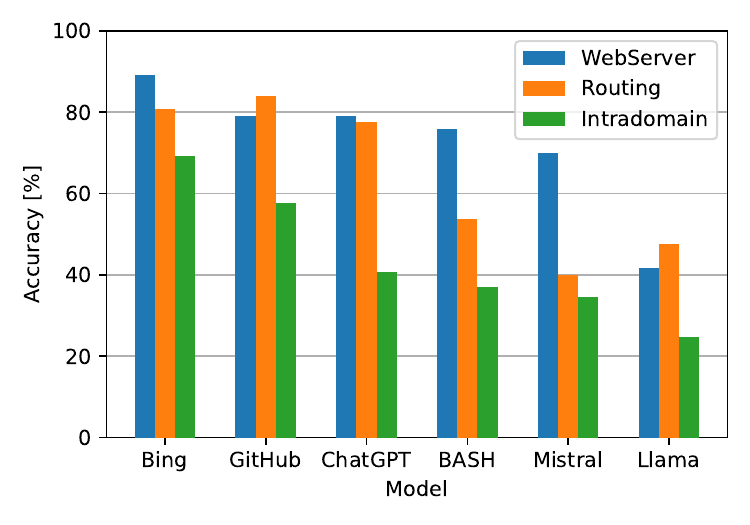}
    \caption{Average accuracy on answering questions on each network.}
    \label{fig:hist}
\end{figure}

As shown in Figure~\ref{fig:hist},
results are variable based on both the model and the network employed.
While proprietary models maintain a discrete accuracy over all the networks, local models show lower results even in simple networks and a more enhanced performance decrease while increasing network complexity. 
Bing is the best-performing model, reaching 8.9 on the most straightforward networks and 6.9 on the most difficult ones, with an overall mean of 7.9. On the contrary, Llama shows the worst performance, with an overall mean accuracy of 3,7.

For the proprietary \acp{llm}, we use the default values for the parameters for their practicality and alignment with typical operational scenarios.
In particular, we employ the base ChatGPT and Copilot while using the \emph{precise} version of Bing.
ChatGPT has higher variability, which can be seen in the results with more intermediate values with respect to the counterparts. 
For local models, we have instead complete control over the parameters.
We tweak our local model temperature parameters to show a small but present response variability.
We employ a temperature of $1.2$ in a scale from 0 to 2 used by the GPT4All Python APIs~\cite{gpt4all}.
In this scale, temperatures above 1 progressively promote equal consideration of all token candidates, while at a temperature between 0 and 1, the likelihood of the best token candidates grows even more.

In the following, we present the results while discussing the research questions and providing some important takeaways.

\subsection{RQ1: Are \acp{llm} able to correctly answer basic questions on network topologies?}

To address this research question, we generate simple queries to identify easily detectable notions that can be extracted from a network, such as the number of machines (\hyperref[tab:quest_gen]{\texttt{T1}}) and IP addresses (\hyperref[tab:quest_gen]{\texttt{T2}-\texttt{T3}}).
We summarize the results in Table~\ref{tab:size}.

\begin{table}[tbh]
\caption{Accuracy on questions regarding the size of the network, reported for the three different networks\protect\footnotemark.}\label{tab:size}\centering \small
\resizebox{\columnwidth}{!}{
\begin{tabular}{l|ccc|ccc|ccc} \hline
    \multirow{2}{*}{\textbf{\ac{llm}}} & \multicolumn{3}{c|}{\texttt{\textbf{T1}}}                                                     & \multicolumn{3}{c|}{\texttt{\textbf{T2}}}                                                     & \multicolumn{3}{c}{\texttt{\textbf{T3}}}                                                        \\
\textbf{}        & \textbf{W}                 & \textbf{R}                 & \textbf{I}                 & \textbf{W}                 & \textbf{R}                & \textbf{I}                  & \textbf{W}                  & \textbf{R}                  & \textbf{I}                 \\ \hline
\textbf{Bing}    & \cellcolor[HTML]{B6D7A8}10 & \cellcolor[HTML]{B6D7A8}10 & \cellcolor[HTML]{B6D7A8}10 & \cellcolor[HTML]{B6D7A8}10 & \cellcolor[HTML]{F4CCCC}0 & \cellcolor[HTML]{F4CCCC}0   & \cellcolor[HTML]{B6D7A8}10  & \cellcolor[HTML]{B6D7A8}10  & \cellcolor[HTML]{B6D7A8}10 \\
\textbf{Github}  & \cellcolor[HTML]{B6D7A8}\textit{10} & \cellcolor[HTML]{B6D7A8}10 & \cellcolor[HTML]{B6D7A8}10 & \cellcolor[HTML]{B6D7A8}10 & \cellcolor[HTML]{D4E2B7}8 & \cellcolor[HTML]{F4CCCC}\textit{0}   & \cellcolor[HTML]{B6D7A8}10  & \cellcolor[HTML]{EAEAC2}6,5 & \cellcolor[HTML]{B6D7A8}\textit{10} \\
\textbf{ChatGPT} & \cellcolor[HTML]{B6D7A8}10 & \cellcolor[HTML]{B6D7A8}10 & \cellcolor[HTML]{E2E8BE}7  & \cellcolor[HTML]{B6D7A8}10 & \cellcolor[HTML]{FAE2CC}3 & \cellcolor[HTML]{FAE2CC}3   & \cellcolor[HTML]{B6D7A8}10  & \cellcolor[HTML]{B6D7A8}10  & \cellcolor[HTML]{F1EDC5}6  \\ \hline
\textbf{BASH Mistral}    & \cellcolor[HTML]{B6D7A8}10 & \cellcolor[HTML]{B6D7A8}10 & \cellcolor[HTML]{F4CCCC}0  & \cellcolor[HTML]{B6D7A8}10 & \cellcolor[HTML]{F4CCCC}0 & \cellcolor[HTML]{F4CCCC}0   & \cellcolor[HTML]{EAEAC2}6,5 & \cellcolor[HTML]{FFF2CC}5   & \cellcolor[HTML]{F4CCCC}0  \\
\textbf{Mistral} & \cellcolor[HTML]{B6D7A8}10 & \cellcolor[HTML]{F4CCCC}0  & \cellcolor[HTML]{F4CCCC}0  & \cellcolor[HTML]{B6D7A8}10 & \cellcolor[HTML]{F4CCCC}0 & \cellcolor[HTML]{F4CCCC}0   & \cellcolor[HTML]{D4E2B7}8   & \cellcolor[HTML]{F9DFCC}2,5 & \cellcolor[HTML]{F4CCCC}0  \\
\textbf{Llama}   & \cellcolor[HTML]{B6D7A8}10 & \cellcolor[HTML]{B6D7A8}10 & \cellcolor[HTML]{F4CCCC}0  & \cellcolor[HTML]{B6D7A8}10 & \cellcolor[HTML]{F4CCCC}0 & \cellcolor[HTML]{F7D7CC}1,5 & \cellcolor[HTML]{F9DFCC}2,5 & \cellcolor[HTML]{FCEACC}4   & \cellcolor[HTML]{F4CCCC}0 \\ \hline
\multicolumn{10}{l}{\footnotesize{\texttt{W}: \netOne{}. \texttt{R}: \netTwo{}. \texttt{I}: \netThree{}.}}
\end{tabular}}
\end{table}
\footnotetext{In certain cases whose scores are indicated in \textit{italic}, we added \textit{``without writing any code''} to the prompt to force Github Copilot to directly return the answer instead of a code snippet to solve the problem.}

The number of nodes is relatively easy to understand for both a human and a machine, as we can see in the table (\hyperref[tab:quest_gen]{\texttt{T1}}).
While almost all the tested models can correctly count nodes in smaller networks with up to three machines, with increasing network size, models showed a significant loss in performance.
This is especially true for open-source models that ultimately return wrong guesses on the number of nodes for extensive networks.
However, proprietary \acp{llm} retain a discrete success rate, especially those employing GPT-4 that maintain 100\% accuracy.  

On the other hand, the number of IP addresses (\hyperref[tab:quest_gen]{\texttt{T2}}), which is usually more significant than the number of machines, is trickier to identify in large networks.
Performances of huge models such as GPT-4 decrease starting from the \netTwo{} scenario. 
All the models performed poorly when the number of addresses increased significantly in the \netThree{} scenario, where all the \acp{llm} are almost unusable.
Indeed, it is known that \acp{llm} like ChatGPT has problems with math~\cite{frieder2024mathematical}, which makes this issue particularly visible in extensive networks.

The last question (\hyperref[tab:quest_gen]{\texttt{T3}}) investigates partial counting, implicitly asking the network to differentiate between addresses assigned to different nodes.
In this case, the math involved is simple, as the maximum number of addresses per node was 4.
Nonetheless, the steps the \ac{llm} needs to follow to answer the question are strict: \textit{(i)} counting the number of addresses for each node; \textit{(ii)} identify the maximum number(s); and \textit{(iii)} return the name of the associated node(s).
While we notice an overall sufficient result for online models, local models suffer even in smaller networks.
This indicated difficulty in reasoning local models, while bigger online \acp{llm} are more prone to think step by step implicitly~\cite{kojima2022largeStepByStep}.  

\begin{formal}
    \textbf{Takeaway}: \acp{llm} can comprehend small networks and answer questions about their size.
    However, they struggle with calculations and increased network complexity. 
\end{formal}

\subsection{RQ2: Can LLMs provide graphical representations of network topologies?}

Graphical representation of a network is challenging for an \ac{llm} but can also prove challenging for a network specialist.
We investigate this question through prompt \hyperref[tab:quest_gen]{\texttt{T4}}.
Since not all the models can generate images (i.e., only Bing can through DALL-E 3~\cite{betker2023improvingDalle}), we included the possibility of using ASCII art to depict the network~\cite{xu2010structure}.
Even though it could be challenging for extensive networks such as \netThree, we provide the \ac{llm} freedom of choosing the detail level of the representations.

As we can see in Table~\ref{tab:draw}, on overall bad results, the reliability in the drawing seems to be linked to the size of the network to be represented.
For networks of a couple of nodes only, results are generally sufficient.
With three nodes and five subnetworks of \netTwo{}, not all the LLMs can provide a representation of the system.
As expected, increasing the complexity of the network again dramatically decreases the results.
No \ac{llm} could represent \netThree{} successfully.
A strategy sometimes adopted by ChatGPT is removing addresses from the ASCII art, which can benefit such extensive networks.
However, it misinterprets node connections, resulting in the wrong schema.

\begin{table}[htb]
\caption{Accuracy on drawing the network on the three scenarios.}\label{tab:draw}\centering \small
\begin{tabular}{l|ccc} \hline
                      \multirow{2}{*}{\textbf{LLM}} & \multicolumn{3}{c}{\texttt{\textbf{T4}}}                                                       \\
                      & \textbf{W}                  & \textbf{R}                  & \textbf{I}                \\ \hline
\textbf{Bing}         & \cellcolor[HTML]{F4CCCC}0   & \cellcolor[HTML]{FAE2CC}3   & \cellcolor[HTML]{F4CCCC}0 \\
\textbf{Github}       & \cellcolor[HTML]{FFF2CC}5   & \cellcolor[HTML]{FFF2CC}5   & \cellcolor[HTML]{F6D3CC}1 \\
\textbf{ChatGPT}      & \cellcolor[HTML]{C5DDB0}9   & \cellcolor[HTML]{FAE2CC}3   & \cellcolor[HTML]{F4CCCC}0 \\ \hline
\textbf{BASH Mistral} & \cellcolor[HTML]{EAEAC2}6,5 & \cellcolor[HTML]{F8DBCC}2   & \cellcolor[HTML]{F4CCCC}0 \\
\textbf{Mistral}      & \cellcolor[HTML]{EAEAC2}6,5 & \cellcolor[HTML]{F7D7CC}1,5 & \cellcolor[HTML]{F6D3CC}1 \\
\textbf{Llama}        & \cellcolor[HTML]{F6D3CC}1   & \cellcolor[HTML]{F4CCCC}0   & \cellcolor[HTML]{F4CCCC}0 \\ \hline
\end{tabular}
\end{table}

\begin{figure}[b]
    \centering
    \includegraphics[width=.75\columnwidth]{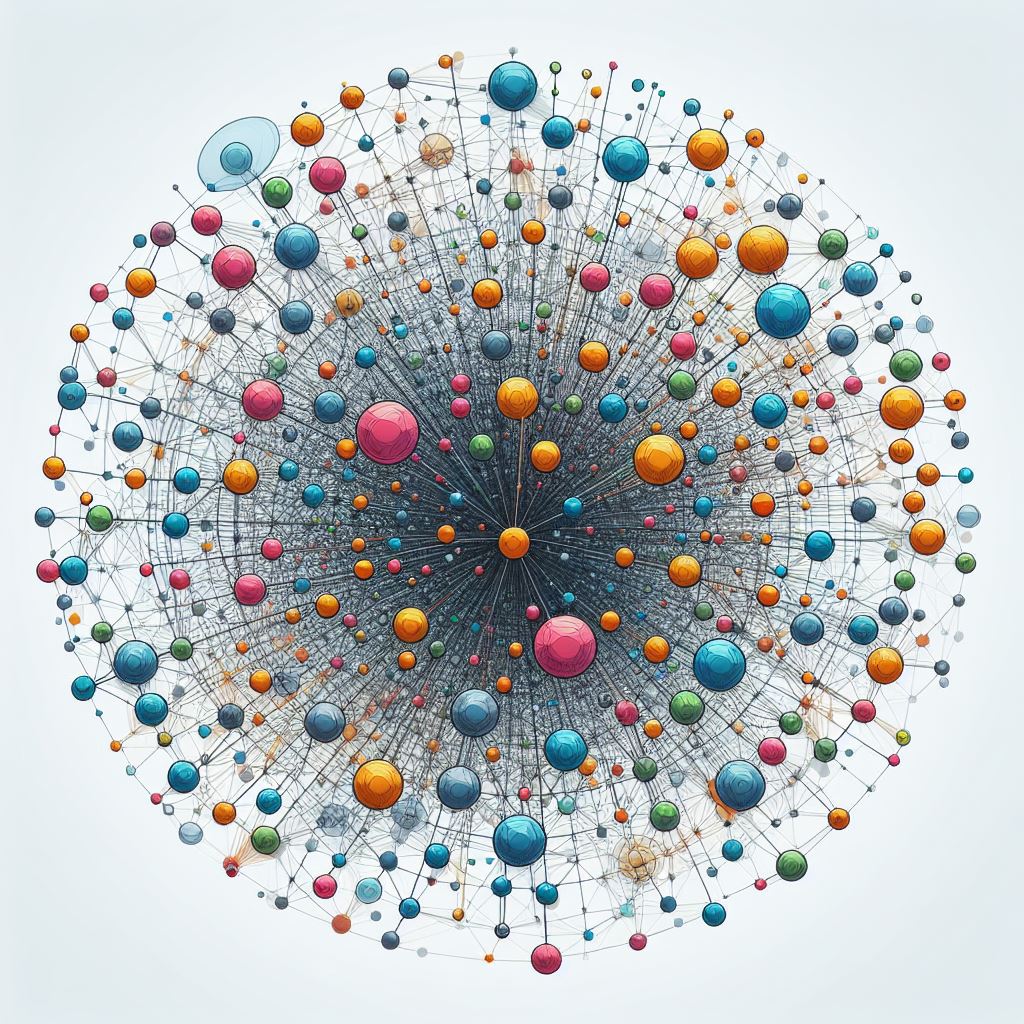}
    \caption{DALL-E generated image of a network. As shown, it is an artistic interpretation and does not include any detail of the original network.}
    \label{fig:dalle-net}
\end{figure}

Bing generates different results, enhanced through DALL-E 3~\cite{betker2023improvingDalle}, a powerful text-to-image \ac{llm}.
Because of this feature, Bing almost always relies on it to draw the schema.
However, generated images are an artistic representation of networks without resembling the network in the prompt.
An example is shown in Figure~\ref{fig:dalle-net}.

\begin{formal}
    \textbf{Takeaway}: \acp{llm} struggle in representing, in any form, network graphs of not-trivial computer networks without including a lot of inaccuracies. 
\end{formal}

\subsection{RQ3: Can \acp{llm} recognize subnetworks and IP addresses inside computer networks?}\label{subsec:ip}

To measure the capabilities of recognizing IP addresses in the context of computer networks, we designed three questions (\hyperref[tab:quest_gen]{\texttt{T5}}, \hyperref[tab:quest_gen]{\texttt{T6}}, and \hyperref[tab:quest_gen]{\texttt{T7}}).
As shown in Table~\ref{tab:addr}, except for Llama, all the networks obtained significant results in these tasks.
Question \hyperref[tab:quest_gen]{\texttt{T5}} discussed the capabilities of individuating specific addresses with meanings that differentiate them from other standard addresses.
The questions include an example to guide the \acp{llm} and reduce hallucination.
ChatGPT responses, on the other hand, contained incorrect answers (e.g., stating that normal addresses have multicast special purposes) and negations of the same answers within the generated text, as follows:
\begin{LLMsnippet}%
\footnotesize
    {
    \fontfamily{qcr}\selectfont
    ``[...] device \addr{as100r1} has an IP address of \addr{140.0.0.2/30}, which falls within the reserved address space for multicast addresses. Multicast addresses typically fall within the range of \addr{224.0.0.0} to \addr{239.255.255.255}. Therefore, \addr{140.0.0.2/30} is within this range and can be considered a multicast address.''
    }
\end{LLMsnippet}
This highlights the difficulties \acp{llm} have in correctly reasoning on the answer they provide.
However, more complex networks such as GPT-4 can answer such questions correctly.

Separating the number of addresses from the devices to which they were assigned was evaluated by question \hyperref[tab:quest_gen]{\texttt{T6}}.
As reported, except for the weird results of Llama, all the other models performed flawlessly, even in more complex networks, showing a good understanding of the separation between devices in the provided NetJSON data.
Similarly, \hyperref[tab:quest_gen]{\texttt{T7}} also reported outstanding results, but the Llama model still presents similar issues to the other questions.
In this task, we measured the capabilities of differentiating IPv4 and IPv6 addresses, a straightforward task for an engineer.

\begin{formal}
    \textbf{Takeaway}: Even inside complex networks, \acp{llm} can recognize different IP address types and properly assign them to the corresponding machine. 
\end{formal}

\begin{table}[htb]
\caption{Accuracy on questions regarding IP addressing, reported for the three different networks.}\label{tab:addr}\centering \small
\resizebox{\columnwidth}{!}{
\begin{tabular}{l|ccc|ccc|ccc} \hline
                      \multirow{2}{*}{\textbf{LLM}} & \multicolumn{3}{c|}{\texttt{\textbf{T5}}}                                                     & \multicolumn{3}{c|}{\texttt{\textbf{T6}}}                                                      & \multicolumn{3}{c}{\texttt{\textbf{T7}}}                                                      \\
                      & \textbf{W}                 & \textbf{R}                 & \textbf{I}                 & \textbf{W}                 & \textbf{R}                 & \textbf{I}                  & \textbf{W}                 & \textbf{R}                 & \textbf{I}                 \\ \hline
\textbf{Bing}         & \cellcolor[HTML]{B6D7A8}10 & \cellcolor[HTML]{B6D7A8}10 & \cellcolor[HTML]{B6D7A8}10 & \cellcolor[HTML]{B6D7A8}10 & \cellcolor[HTML]{B6D7A8}10 & \cellcolor[HTML]{B6D7A8}10  & \cellcolor[HTML]{B6D7A8}10 & \cellcolor[HTML]{B6D7A8}10 & \cellcolor[HTML]{B6D7A8}10 \\
\textbf{Github}       & \cellcolor[HTML]{B6D7A8}10 & \cellcolor[HTML]{B6D7A8}\textit{10} & \cellcolor[HTML]{B6D7A8}\textit{10} & \cellcolor[HTML]{B6D7A8}10 & \cellcolor[HTML]{B6D7A8}10 & \cellcolor[HTML]{B6D7A8}10  & \cellcolor[HTML]{B6D7A8}10 & \cellcolor[HTML]{B6D7A8}10 & \cellcolor[HTML]{B6D7A8}10 \\
\textbf{ChatGPT}      & \cellcolor[HTML]{B6D7A8}10 & \cellcolor[HTML]{B6D7A8}10 & \cellcolor[HTML]{FAE2CC}3  & \cellcolor[HTML]{B6D7A8}10 & \cellcolor[HTML]{B6D7A8}10 & \cellcolor[HTML]{B6D7A8}10  & \cellcolor[HTML]{B6D7A8}10 & \cellcolor[HTML]{B6D7A8}10 & \cellcolor[HTML]{B6D7A8}10 \\ \hline
\textbf{BASH Mistral} & \cellcolor[HTML]{B6D7A8}10 & \cellcolor[HTML]{B6D7A8}10 & \cellcolor[HTML]{B6D7A8}10 & \cellcolor[HTML]{B6D7A8}10 & \cellcolor[HTML]{B6D7A8}10 & \cellcolor[HTML]{B6D7A8}10  & \cellcolor[HTML]{B6D7A8}10 & \cellcolor[HTML]{B6D7A8}10 & \cellcolor[HTML]{B6D7A8}10 \\
\textbf{Mistral}      & \cellcolor[HTML]{B6D7A8}10  & \cellcolor[HTML]{B6D7A8}10 & \cellcolor[HTML]{B6D7A8}10 & \cellcolor[HTML]{B6D7A8}10 & \cellcolor[HTML]{B6D7A8}10 & \cellcolor[HTML]{B6D7A8}10  & \cellcolor[HTML]{B6D7A8}10 & \cellcolor[HTML]{B6D7A8}10 & \cellcolor[HTML]{B6D7A8}10 \\
\textbf{Llama}        & \cellcolor[HTML]{C5DDB0}9  & \cellcolor[HTML]{F1EDC5}6  & \cellcolor[HTML]{FFF2CC}5  & \cellcolor[HTML]{FCEACC}4  & \cellcolor[HTML]{E2E8BE}7  & \cellcolor[HTML]{F9DFCC}2,5 & \cellcolor[HTML]{FCEACC}4  & \cellcolor[HTML]{FFF2CC}5  & \cellcolor[HTML]{B6D7A8}10 \\ \hline
\end{tabular}}
\end{table}

While comprehending IP addresses has proven to be usually straightforward for \acp{llm}, they perform far worse in recognizing subnets, as shown in Table~\ref{tab:sub}.
Question \hyperref[tab:quest_gen]{\texttt{T8}} investigated the capabilities of the \acp{llm} to compute and count subnetworks.
If it could be a time-demanding yet straightforward task for a network engineer, it is more difficult for an \ac{llm}, which can provide wrong answers even with simple networks. 
For example, we can see a wrong reasoning from ChatGPT-3.5, stating that two IP addresses belong to two different subnetworks.
However, it is not true with the \texttt{/16} netmask: 
\begin{LLMsnippet}\footnotesize
    {
    \fontfamily{qcr}\selectfont
    ``In the given network, both devices [...] belong to separate subnetworks \addr{(10.0.0.0/16} and \addr{10.0.1.0/16}).''
    }
\end{LLMsnippet} 
The error probably originated from the structure of the employed IP addresses that resemble two different \texttt{/24} subnetworks.
However, this is a clear example showing how the \ac{llm} cannot comprehend the subnetting task correctly but instead bases its reasoning on the form of the training data. 

Github Copilot, developed to write code, tries instead to answer with a script and often prefers not to answer at all when explicitly asked not to do so.
Overall, the main problem related to this task was complete ignorance of the netmask by many answers.
Finally, it is worth mentioning the good results of BASH Mistral on \netOne, which was probably due to the training set and not superior performances, as confirmed by the following networks. 

The last question (\hyperref[tab:quest_gen]{\texttt{T9}}) contains more reasoning connected to the network structure instead.
In particular, it requires different steps, such as computing subnetworks and checking device connections.
In this task, results are scattered, with a high predominance of Bing, which solved the task on all the networks.
Moreover, it provides interesting insights showing the capabilities of considering both IPv4 and IPv6 addresses: 

\begin{LLMsnippet}\footnotesize
    {
    \fontfamily{qcr}\selectfont{
    ``If you remove the IPv4 subnetwork (\addr{10.0.0.0/16} and \addr{10.0.1.0/16}), the nodes will still be able to communicate with each other using their IPv6 addresses, assuming that your network infrastructure supports IPv6.''}
    }
\end{LLMsnippet}
Other models, instead, have more difficulties on this task. ChatGPT, for instance, was not able to assess the presence of a unique IPv4 subnet in network \netOne:
\begin{LLMsnippet}\footnotesize     
{
\fontfamily{qcr}\selectfont{
``[...] In the given network, there are two subnetworks: %
    (1) Subnetwork with IP range \addr{10.0.0.0/16} (containing the \addr{client}) %
    (2) Subnetwork with IP range \addr{10.0.1.0/16} (containing the \addr{server})''}
    }
\end{LLMsnippet}
When the complexity of the network increased, Github and ChatGPT started providing generic answers unrelated to the proposed network or saying that they could not answer the question, respectively.

\begin{table}[htb]
\caption{Accuracy on subnetting questions on the three scenarios.}\label{tab:sub}\centering \small
\resizebox{\columnwidth}{!}{
\begin{tabular}{l|ccc|ccc|ccc}
\hline
                               & \multicolumn{3}{c|}{\texttt{\textbf{T8}}}                                                                     & \multicolumn{3}{c|}{\texttt{\textbf{T9}}}                                                      & \multicolumn{3}{c}{\texttt{\textbf{T10}}}                                                     \\
\multirow{-2}{*}{\textbf{LLM}} & \textbf{W}                 & \textbf{R}                         & \textbf{I}                         & \textbf{W}                 & \textbf{R}                 & \textbf{I}                  & \textbf{W}                 & \textbf{R}                 & \textbf{I}                 \\ \hline
\textbf{Bing}                  & \cellcolor[HTML]{F8DBCC}2  & \cellcolor[HTML]{F8DBCC}2          & \cellcolor[HTML]{D4E2B7}8          & \cellcolor[HTML]{B6D7A8}10 & \cellcolor[HTML]{B6D7A8}10 & \cellcolor[HTML]{B6D7A8}10  & \cellcolor[HTML]{B6D7A8}10 & \cellcolor[HTML]{B6D7A8}10     & \cellcolor[HTML]{B6D7A8}10 \\
\textbf{Github}                & \cellcolor[HTML]{F4CCCC}0  & \cellcolor[HTML]{F4CCCC}\textit{0} & \cellcolor[HTML]{F4CCCC}\textit{0} & \cellcolor[HTML]{B6D7A8}10 & \cellcolor[HTML]{B6D7A8}10 & \cellcolor[HTML]{F4CCCC}0   & \cellcolor[HTML]{B6D7A8}10 & \cellcolor[HTML]{B6D7A8}10     & \cellcolor[HTML]{B6D7A8}10 \\
\textbf{ChatGPT}               & \cellcolor[HTML]{F6D3CC}1  & \cellcolor[HTML]{FCEACC}4          & \cellcolor[HTML]{F4CCCC}0          & \cellcolor[HTML]{F4CCCC}0  & \cellcolor[HTML]{E2E8BE}7  & \cellcolor[HTML]{FBE6CC}3,5 & \cellcolor[HTML]{B6D7A8}10 & \cellcolor[HTML]{B6D7A8}10     & \cellcolor[HTML]{F8DBCC}2  \\ \hline
\textbf{BASH Mistral}          & \cellcolor[HTML]{B6D7A8}10 & \cellcolor[HTML]{F8DBCC}2          & \cellcolor[HTML]{F4CCCC}0          & \cellcolor[HTML]{FCEACC}4  & \cellcolor[HTML]{D4E2B7}8  & \cellcolor[HTML]{D4E2B7}8   & \cellcolor[HTML]{B6D7A8}10 & \cellcolor[HTML]{F4CCCC}0      & \cellcolor[HTML]{F4CCCC}0  \\
\textbf{Mistral}               & \cellcolor[HTML]{F4CCCC}0  & \cellcolor[HTML]{F4CCCC}0          & \cellcolor[HTML]{F4CCCC}0          & \cellcolor[HTML]{C5DDB0}9  & \cellcolor[HTML]{F4CCCC}0  & \cellcolor[HTML]{F5CFCC}0,5 & \cellcolor[HTML]{C5DDB0}9  & \cellcolor[HTML]{F4CCCC}0      & \cellcolor[HTML]{F9DFCC}2,5  \\
\textbf{Llama}                 & \cellcolor[HTML]{F4CCCC}0  & \cellcolor[HTML]{FFF2CC}5          & \cellcolor[HTML]{F4CCCC}0          & \cellcolor[HTML]{F8DBCC}2  & \cellcolor[HTML]{FAE2CC}3  & \cellcolor[HTML]{F6D3CC}1   & \cellcolor[HTML]{F4CCCC}0  & \cellcolor[HTML]{B6D7A8}10     & \cellcolor[HTML]{F4CCCC}0  \\ \hline
\end{tabular}%
}%

\end{table}

Then, we check if \acp{llm} can identify the subnetwork connection of two machines equipped with at least two different addresses each (\hyperref[tab:quest_gen]{\texttt{T10}}).
While online models usually return perfect results, offline models generally perform worse, apart from some spikes.
As expected, results are worse for the \netThree{} network.
In particular, we notice how, in the presence of many links between nodes, big models are prone to make up connections that do not exist in the topology, thus leading to wrong final answers.

\begin{formal}
    \textbf{Takeaway:} Proprietary LLMs can provide answers regarding subnetting only if simple tasks are required. Instead, open-source LLMs suffer from unsteady results and, up to now, are unreliable. 
\end{formal}

\subsection{RQ4: Can LLMs comprehend computer network connections?}

In this section, we analyze the capabilities of \acp{llm} in path computation between nodes inside each network.
Results are summarized in Table~\ref{tab:path_comp}.

\begin{table}[htb]
\caption{Accuracy on path computation questions on the three scenarios.}\label{tab:path_comp}\centering \small
\resizebox{\columnwidth}{!}{
\begin{tabular}{l|ccc|ccc|ccc}
\hline
                               & \multicolumn{3}{c|}{\texttt{\textbf{T11}}}                                                                     & \multicolumn{3}{c|}{\texttt{\textbf{T12}}}                                                   & \multicolumn{3}{c}{\texttt{\textbf{T13}}}                                                        \\
\multirow{-2}{*}{\textbf{LLM}} & \textbf{W\protect\footnotemark}                & \textbf{R}                          & \textbf{I}                          & \textbf{W}                 & \textbf{R}                 & \textbf{I}                & \textbf{W}                  & \textbf{R}                  & \textbf{I}                  \\ \hline
\textbf{Bing}                  & \cellcolor[HTML]{D9D9D9}- & \cellcolor[HTML]{B6D7A8}10          & \cellcolor[HTML]{B6D7A8}10          & \cellcolor[HTML]{B6D7A8}10 & \cellcolor[HTML]{B6D7A8}10 & \cellcolor[HTML]{F4CCCC}0 & \cellcolor[HTML]{B6D7A8}10  & \cellcolor[HTML]{B6D7A8}10  & \cellcolor[HTML]{F7D7CC}1,5 \\
\textbf{Github}                & \cellcolor[HTML]{D9D9D9}- & \cellcolor[HTML]{B6D7A8}\textit{10} & \cellcolor[HTML]{B6D7A8}\textit{10} & \cellcolor[HTML]{B6D7A8}10 & \cellcolor[HTML]{B6D7A8}10 & \cellcolor[HTML]{F4CCCC}0 & \cellcolor[HTML]{F4CCCC}0   & \cellcolor[HTML]{CCE0B3}8,5 & \cellcolor[HTML]{FCEACC}4   \\
\textbf{ChatGPT}               & \cellcolor[HTML]{D9D9D9}- & \cellcolor[HTML]{B6D7A8}10          & \cellcolor[HTML]{FAE2CC}3           & \cellcolor[HTML]{C5DDB0}9  & \cellcolor[HTML]{F1EDC5}6  & \cellcolor[HTML]{FAE2CC}3 & \cellcolor[HTML]{F8F0C9}5,5 & \cellcolor[HTML]{DBE5BA}7,5 & \cellcolor[HTML]{F8DBCC}2   \\ \hline
\textbf{BASH Mistral}          & \cellcolor[HTML]{D9D9D9}- & \cellcolor[HTML]{B6D7A8}10          & \cellcolor[HTML]{B6D7A8}10          & \cellcolor[HTML]{F4CCCC}0  & \cellcolor[HTML]{F4CCCC}0  & \cellcolor[HTML]{F4CCCC}0 & \cellcolor[HTML]{FAE2CC}3   & \cellcolor[HTML]{F9DFCC}2,5 & \cellcolor[HTML]{F4CCCC}0   \\
\textbf{Mistral}               & \cellcolor[HTML]{D9D9D9}- & \cellcolor[HTML]{B6D7A8}10          & \cellcolor[HTML]{B6D7A8}10          & \cellcolor[HTML]{F4CCCC}0  & \cellcolor[HTML]{F4CCCC}0  & \cellcolor[HTML]{F4CCCC}0 & \cellcolor[HTML]{F6D3CC}1   & \cellcolor[HTML]{E2E8BE}7   & \cellcolor[HTML]{F4CCCC}0   \\
\textbf{Llama}                 & \cellcolor[HTML]{D9D9D9}- & \cellcolor[HTML]{FFF2CC}5           & \cellcolor[HTML]{F1EDC5}6           & \cellcolor[HTML]{FCEACC}4  & \cellcolor[HTML]{F8DBCC}2  & \cellcolor[HTML]{FFF2CC}5 & \cellcolor[HTML]{F9DFCC}2,5 & \cellcolor[HTML]{FFF2CC}5   & \cellcolor[HTML]{F4CCCC}0   \\ \hline
\end{tabular}%
}%
\end{table}
\footnotetext{Since \netOne{} contains two directly connected nodes only, this question is the same as \hyperref[tab:quest_gen]{\texttt{T12}} and requires a positive answer.}

We started with the more straightforward tasks asking the \acp{llm} to test the capabilities of directly ping one machine from another without any hop (\hyperref[tab:quest_gen]{\texttt{T11}} and \hyperref[tab:quest_gen]{\texttt{T12}}).
While \hyperref[tab:quest_gen]{\texttt{T11}} seems to return high scores, it may be due to a bias of \acp{llm} to answer negatively when they are not sure about the correct final answer.
This is verifiable with questions \hyperref[tab:quest_gen]{\texttt{T12}}, where the requested connections were instead possible without any hop.
Here, only GPT4-based models maintain similar scores, while the other models generally reduce their accuracy.
This highlights how the models could not correctly understand the question and the reasoning steps to compute a correct answer. 

Bad results are also shown in the \netThree{} scenario, with even lower performances for all the tasks.
Even GPT-4-based models cannot correctly answer questions on such an extensive network.

Moreover, we try to see how \acp{llm} work when asked to return all the hops between two nodes (\hyperref[tab:quest_gen]{\texttt{T13}}).
The terminal command to get that information is called \texttt{traceroute}, and we trust the \acp{llm} to have been trained with data containing that command.
Surprisingly, the Mistral fine-tuned on BASH commands performed poorly, generally worse than other not-fine-tuned offline models, demonstrating not to have grasped the meaning of the command.
Instead, other proprietary \acp{llm} showed discrete results in \netTwo{}, while scores drastically fell for the biggest \netThree{}.
Moreover, with the simple scenario \netOne{}, GitHub returns only the command that needs to be executed without following the link and returning the result.  
This shows the difficulties for all the \acp{llm} to perform inherently multi-stage tasks such as computing the path between two nodes.

\begin{formal}
    \textbf{Takeaway:} While it is generally possible for GPT-4 models to compute paths between nodes, performances drastically reduce with the increase of the network size. 
\end{formal}

\section{Discussion}
\label{sec:discussion}

In this section, we discuss the obtained results. Section~\ref{subsec:prompt_eng} investigates the usage of prompt engineering to improve performances, Section~\ref{subsec:analysis} discusses the \ac{llm} reasoning process, while Section~\ref{subsec:privacy} analyses security and privacy concerns.

\subsection{Prompt Engineering}\label{subsec:prompt_eng} %

In this section, we explore the possibility of enhancing results through prompt engineering~\cite{white2023prompt, velasquez2023prompt}.
We focus our effort on the overall best-performing model, which is the GPT-4-powered Bing~\cite{bing}.
As suggested in the literature, some enhancements can be added to the base prompt (\texttt{B}, shown in Figure~\ref{fig:prompt}) to improve the results.
In particular, we employed the well-known \textit{``reason step by step''} sentence (\texttt{SR}) to force the \ac{llm} into describing its reasoning steps~\cite{kojima2022largeStepByStep}.
We insert this text at the end of the prompt or, when available, before indicating how to formulate the answer (e.g., \textit{``answer with a number''}). In these cases, we also add \textit{``and then''} to enhance the sequentiality of the tasks. 
Then, we enhance the characterization of the \ac{llm} inserting his role as a senior network engineer, as previously done in a similar contexts~\cite{wang2024wordflow,uusnakki2023design}, inserting the sentence \textit{``you are a senior network engineer''} (\texttt{NE}) at the beginning of the prompt. 
Finally, we try an improvement we designed by looking at previous mistakes in the traceroute experiment (\hyperref[tab:quest_gen]{\texttt{T13}}) in the \netThree{} scenario.
Since mistakes are usually due to the system making up connections, we ask the \ac{llm} to \textit{``explain all the relevant connections between two nodes''} (\texttt{RC}) to force it to reason about it correctly.
We add this at the end of the prompt.
The results are summarized in Table~\ref{tab:prompt_eng}.

\begin{table}[htb]
\caption{Results using prompt engineering techniques with the Bing model.
B: Base prompt; SR: ``Reason step by step"; NE: ``You are a senior network engineer."; RC: ``Explain all the relevant connections between two nodes.``; All: all the above.
}\label{tab:prompt_eng} \centering \small
\begin{tabular}{c|c|c|ccc|c}
\hline
                                          \multirow{2}{*}{\textbf{Task}} & \multirow{2}{*}{\textbf{Network}} & \multirow{2}{*}{\textbf{B}} & \multicolumn{4}{c}{\textbf{Prompt}}                                                                                                           \\ \cline{4-7}
& &                  & \textbf{\texttt{SR}}                & \textbf{\texttt{NE}}               & \textbf{\texttt{RC}}                & \textbf{All}              \\ \hline
\multirow{2}{*}{\texttt{T2}} & \netTwo{} (R)                           & \cellcolor[HTML]{F4CCCC}0   & \cellcolor[HTML]{B6D7A8}10  & \cellcolor[HTML]{F4CCCC}0 & \cellcolor[HTML]{FCEACC}4  & \cellcolor[HTML]{E2E8BE}7 \\
 & \netThree{} (I)                         & \cellcolor[HTML]{F4CCCC}0   & \cellcolor[HTML]{B6D7A8}10 & \cellcolor[HTML]{F4CCCC}0 & \cellcolor[HTML]{F4CCCC}0  & \cellcolor[HTML]{C5DDB0}9 \\ \hline
\multirow{3}{*}{\texttt{T8}} & \netOne{} (W)                            & \cellcolor[HTML]{F8DBCC}2   & \cellcolor[HTML]{FAE2CC}3  & \cellcolor[HTML]{C5DDB0}9 & \cellcolor[HTML]{B6D7A8}10 & \cellcolor[HTML]{D4E2B7}8 \\
 & \netTwo{} (R)                           & \cellcolor[HTML]{F8DBCC}2   & \cellcolor[HTML]{FCEACC}4  & \cellcolor[HTML]{F4CCCC}0 & \cellcolor[HTML]{B6D7A8}10 & \cellcolor[HTML]{FFF2CC}5 \\
 & \netThree{} (I)                         & \cellcolor[HTML]{D4E2B7}8   & \cellcolor[HTML]{D4E2B7}8  & \cellcolor[HTML]{F4CCCC}0 & \cellcolor[HTML]{F8DBCC}2  & \cellcolor[HTML]{E2E8BE}7 \\ \hline
\texttt{T12} & \netThree{} (I)                         & \cellcolor[HTML]{F4CCCC}0   & \cellcolor[HTML]{D4E2B7}8  & \cellcolor[HTML]{F4CCCC}0 & \cellcolor[HTML]{F4CCCC}0  & \cellcolor[HTML]{D4E2B7}8 \\ \hline
\texttt{T13} & \netThree{} (I)                        & \cellcolor[HTML]{F7D7CC}1,5 & \cellcolor[HTML]{F8DBCC}2  & \cellcolor[HTML]{FAE2CC}3 & \cellcolor[HTML]{D4E2B7}8  & \cellcolor[HTML]{F8DBCC}2 \\ \hline
\end{tabular}%
\end{table}

Interesting results are obtained using the \texttt{SR} and require proper discussion.
We notice how they usually improve for complex networks while being almost unchanged for small networks.
This happens because expanding the reasoning is insufficient when the network is simple but contains some tricky structures.
In our case, in \netOne{}, problems were related to the IPv6 assigned, which were not adequately considered. 
However, forcing the system to reason step by step is usually a good practice to improve the quality of responses (as suggested in previous works~\cite{kojima2022largeStepByStep, aksitov2023rest}) also in the computer network scenario.

Regarding \hyperref[tab:quest_gen]{\texttt{T13}} (i.e., traceroute), the primary problem is that the \ac{llm} is unable to identify network connections correctly, and sometimes it makes up links.
This happens even with the other enhancements to the prompt, where one of the steps is usually broken because of an invented link between two nodes. 
In contrast, different connections and the provided explanation seem satisfying.
A well-written but wrong discussion may be dangerous since it may sound correct to the users, tricking them into accepting an incorrect answer. 

\begin{formal}
     \textbf{Takeaway:} Step by step reasoning generally improves results correctness. However, it may induce sound but wrong answers that network engineers should be aware of. 
\end{formal}

While exploring different prompts in different questions, we also combine them to generate a prompt that can work on all the tasks. %
As we can see, it generally improves the results with respect to the basic prompt.
However, results are not always the best with respect to single prompt improvements.
This makes it trickier to develop a unique prompt for a \ac{llm} designed for network engineers.

\begin{formal}
    \textbf{Takeaway:} %
    It is not trivial to develop a prompt template that works on all the various tasks regarding computer networks.
\end{formal}

To overcome the problem, an efficient solution could be to teach network engineers how to provide better prompts to \acp{llm}.
Indeed, from Table~\ref{tab:prompt_eng}, we can see the logic behind specific results.
For instance, we can see how \texttt{RC} is fundamental for \hyperref[tab:quest_gen]{\texttt{T13}} since the \ac{llm} must understand all the links between devices to perform a traceroute correctly.
Adding other improvements can instead decrease the results.

\begin{formal}
    \textbf{Takeaway:} Network engineers should have basic knowledge of prompt engineering to best tune queries and employ \acp{llm} at their best.
\end{formal}

\subsection{Reasoning Analysis}
\label{subsec:analysis}

One of the most promising approaches of prompt engineering (discussed in Section~\ref{subsec:prompt_eng}) is the step by step reasoning, which is an application of the \ac{cot}~\cite{wei2022chain} prompting. It is a recent advancement in prompting methods that encourages \ac{llm} to explain their reasoning. 

\begin{figure}[t]
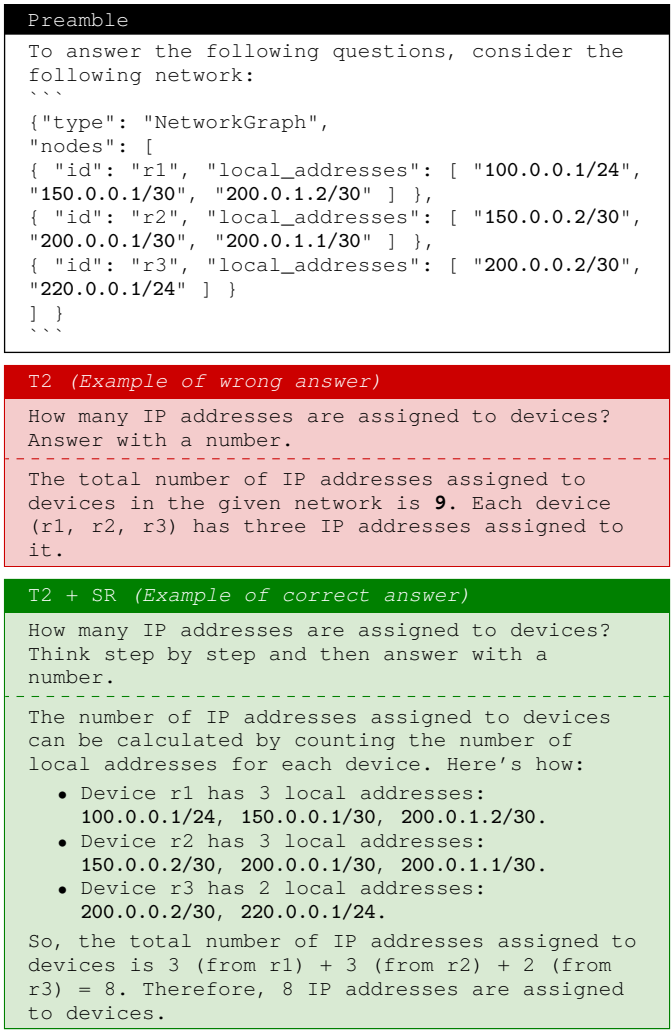


\footnotesize
{
\fontfamily{qcr}\selectfont{
\begin{tcolorbox}[enhanced jigsaw, interior hidden, breakable, sharp corners, colback=white, colframe=black, size=title, boxrule=0.5pt, title=Preamble]
To answer the following questions, consider the following network: \\
\`{}\`{}\`{}\\
\{"type": "NetworkGraph", \\
    "nodes": [ \\ 
            \{ 
            "id": "r1",
            "local\_addresses": [
                "\addr{100.0.0.1/24}",
                "\addr{150.0.0.1/30}",
                "\addr{200.0.1.2/30}"
            ] \}, \\
            \{
            "id": "r2",
            "local\_addresses": [
                "\addr{150.0.0.2/30}",
                "\addr{200.0.0.1/30}",
                "\addr{200.0.1.1/30}"
            ]
        \},  \\
        \{
            "id": "r3",
            "local\_addresses": [
                "\addr{200.0.0.2/30}",
                "\addr{220.0.0.1/24}"
            ]
        \} \\
    ] %
\}\\ 
\`{}\`{}\`{}
\end{tcolorbox}\begin{tcolorbox}[enhanced jigsaw,breakable, sharp corners, colframe=red!80!black, colback=redBello, size=title, boxrule=0.5pt, title=T2 \textit{(Example of wrong answer)}]
How many IP addresses are assigned to devices? Answer with a number.
\tcblower
The total number of IP addresses assigned to devices in the given network is \textbf{9}. Each device (r1, r2, r3) has three IP addresses assigned to it.
\end{tcolorbox}\begin{tcolorbox}[enhanced jigsaw,breakable, sharp corners, colframe=green!50!black, colback=greenBello, size=title, boxrule=0.5pt, title=T2 + SR \textit{(Example of correct answer)}]
How many IP addresses are assigned to devices? Think step by step and then answer with a number.
\tcblower
The number of IP addresses assigned to devices can be calculated by counting the number of local addresses for each device. Here’s how:
\begin{itemize}
    \item Device r1 has 3 local addresses: \addr{100.0.0.1/24}, \addr{150.0.0.1/30}, \addr{200.0.1.2/30}.%
    \item Device r2 has 3 local addresses: \addr{150.0.0.2/30}, \addr{200.0.0.1/30}, \addr{200.0.1.1/30}.%
    \item Device r3 has 2 local addresses: \addr{200.0.0.2/30}, \addr{220.0.0.1/24}.
\end{itemize}
So, the total number of IP addresses assigned to devices is 3 (from r1) + 3 (from r2) + 2 (from r3) = 8. Therefore, 8 IP addresses are assigned to devices.
\end{tcolorbox}
} %
} %
\caption{An example of a complete prompt related to task \texttt{T2} and network \netTwo{}, both with base prompt and with step by step reasoning.}\label{fig:example}
\end{figure}

In Figure~\ref{fig:example} we provide an end-to-end example of how \ac{cot} is capable of fixing wrong reasoning on the best-performing model (Bing). 
Indeed, in the wrong answer, the \ac{llm} simply assigns three addresses to each node. Instead, with \ac{cot}, the \ac{llm} is forced into explicitly counting the addresses in each node, leading to a correct answer.

\subsection{Security and Privacy Aspects}
\label{subsec:privacy}
Deploying \acp{llm} in system administration necessitates a comprehensive examination of security and privacy considerations. 
We delve into two critical aspects: protecting sensitive information and the strategic role of \acp{llm} as supportive tools for system administrators.
It is fundamental to consider security and privacy themes that concern the utilization of \acp{llm} for system administrator operations~\cite{peris2023privacy}. 

First, outsourcing companies' information, including network topology details and connected device types, may not be practically achievable.
Such information often harbors sensitive data, and any accidental leakage could potentially jeopardize the company's security.
Therefore, we believe virtual system administrators should integrate \acp{llm} that can run locally in the company perimeter. 

Second, system administrators frequently encounter requests to execute critical tasks to ensure the network's overall health.
However, relying on \acp{llm} as a black-box solution can lead to the accidental introduction of faulty configurations, potentially jeopardizing the stability and security of the network infrastructure.
Therefore, as transparent \ac{ml} applications are essential in cyber security~\cite{nadeem2023sok}, we believe that technologies such as \acp{llm} can be utilized as a system admin co-pilot (in support) rather than replacing them. 

\section{Related Works}
\label{sec:related}

\acp{llm} have been applied in different fields recently.
They are potent models trained on vast corpora of documents that can achieve general-purpose language generation.
They can find applications in many sectors, such as medicine~\cite{thirunavukarasu2023large}, education~\cite{latif2023knowledge}, coding~\cite{gu2023llm} or cybersecurity~\cite{fang2024llm}.
Other possible tasks where \acp{llm} can improve or simplify solutions are network topology optimization~\cite{gupta2006topology, shakkottai2008network} or network traffic prediction~\cite{oliveira2016computer} and management~\cite{fadlullah2017state, mani2023enhancing}. The telecom domain is going to benefit from \acp{llm} as well~\cite{maatouk2023large, bariah2024large}. For instance, a fine-tuned \ac{llm} can help understanding the telecom language~\cite{bariah2023understanding, ahmed2024linguistic}.
Moreover, %
network understanding could, for example, enhance the usage of \acp{llm} in penetration testing and red teaming tasks~\cite{genevey2023red, happe2023getting}.

Graph reasoning related to \acp{llm} has been extensively studied~\cite{chai2023graphllm, zhang2023graph}.
It describes a cognitive process involving analyzing and interpreting information presented in a graphical form (e.g., charts, diagrams, networks).
It often infers, draws conclusions, or solves problems based on relationships or patterns. 
The other way around, graph-to-text, have been investigated in the literature~\cite{yuan2023evaluating}.
Plot to text has also been researched, including a reasoning step on the generated content~\cite{liu2022deplot}.
While \ac{ai} has been used to perform research on graph networks, no one ever tries to understand the capabilities of off-the-shelf \acp{llm} in understanding computer network topologies. 

\section{Conclusions}\label{sec:conclusions}

This work represents the first step in assessing the capabilities of \acp{llm} in understanding computer networks and answering questions regarding them.
We developed a framework to answer our research questions and establish the reasoning capabilities of \acp{llm} in this field.
From our first analysis, we discover a good accuracy of huge proprietary models in correctly answering questions on simple and complex computer network topologies.
In particular, \acp{llm} can comprehend the topology of smaller networks and correctly recognize and assign IPs to different machines.
However, we highlight several limitations system administrators should know when using \acp{llm} as part of their job.
For example, proprietary \acp{llm} currently yield the best results at the cost of network confidentiality.
Moreover, fine-tuning and prompt engineering may be necessary to make \acp{llm} employable by system administrators in the wild.

\emph{Future Works.}
This paper does not aim to be a comprehensive benchmark of models and parameters in the computer network context but a first exploratory study to understand the overall capabilities of state-of-the-art \acp{llm} in this field.
We foresee several future works in this area.
More models, including models fine-tuned on computer network tasks, should be evaluated using our framework to identify those that perform best in this area.
The variation of accuracy as model parameters (e.g., temperature) change and the format of the prompt ---also including few-shot strategies~\cite{wang2020generalizing}--- should be investigated as well.
Furthermore, on the network side, testing other representation techniques could open more opportunities, as our current approach uses IP addresses to infer physical connections.

\section*{Acknowledgment}
We would like to thank Omitech S.r.l. for supporting the research of Denis Donadel.

\balance
\bibliographystyle{ieeetr}
\bibliography{bibliography}

\end{document}